\pdfoutput=1
\documentclass{PoS}

\usepackage{amsfonts,amsmath,bm}
\usepackage{placeins}

\newcommand{\ua}{\uparrow}
\newcommand{\da}{\downarrow}
\newcommand{\bs}[1]{\ensuremath{{\boldsymbol{#1}}}}

\title{$\Omega_{bbb}$ excited-state spectroscopy from lattice QCD}

\ShortTitle{$\Omega_{bbb}$ excited-state spectroscopy from lattice QCD}

\author{\speaker{Stefan Meinel}\\
Center for Theoretical Physics,\\
Massachusetts Institute of Technology,\\
Cambridge, MA 02139, USA\\
E-mail: \email{smeinel@mit.edu}}

\abstract{
Triply heavy baryons are very interesting systems analogous to heavy quarkonia, but are difficult to access experimentally.
Lattice QCD can provide precise predictions for these systems, which can be compared to other theoretical approaches.
In this work, the spectrum of excited states of the $\Omega_{bbb}$ baryon is calculated using lattice NRQCD for the $b$
quarks, and using a domain-wall action for the $u$, $d$ and $s$ sea quarks. The calculations are done for multiple values
of the sea-quark masses, and for two different lattice spacings. The energies of states with angular momentum up to
$J=7/2$ are calculated, and the effects of rotational symmetry breaking by the lattice are analyzed. Precise results are
obtained even for the small spin-dependent energy splittings, and the contributions of individual NRQCD interactions to
these energy splittings are studied. The results are compared to potential-model calculations.}

\FullConference{Xth Quark Confinement and the Hadron Spectrum,\\
October 8-12, 2012\\
TUM Campus Garching, Munich, Germany}

\begin{document}

\FloatBarrier
\section{Introduction}
\FloatBarrier

The $bbb$ system can be viewed as the baryonic analogue of the bottomonium system. Like bottomonia, $bbb$ baryons are
governed by multiple well-separated energy scales and are therefore amenable to the description with effective field
theories \cite{Brambilla:2005yk}. Baryons exhibit the $SU(3)$ gauge symmetry of QCD more directly than mesons, and are
sensitive to the resulting genuine three-body forces. Triply heavy baryons probe the three-quark interactions at
relatively short distances, where contact with perturbation theory can be made \cite{Brambilla:2009cd}.

The aim of the work reported here is to complement the perturbative QCD calculations and other theoretical studies
of triply heavy baryons with nonperturbative lattice QCD calculations. In Ref.~\cite{Meinel:2010pw}, the mass of the
ground-state $\Omega_{bbb}$ baryon was calculated using lattice QCD to be
$14.371 \pm 0.004_{\rm \:stat} \pm 0.011_{\rm \:syst} \pm 0.001_{\rm \:exp}$ GeV. This was followed by a lattice QCD
calculation of $\Omega_{bbb}$ excited states in Ref.~\cite{Meinel:2012qz}, which is summarized here. Like the bottomonium
spectrum, the $bbb$ spectrum features a hierarchy of radial/orbital excitations and spin-dependent energy splittings.
In Ref.~\cite{Meinel:2012qz}, the energies of ten $bbb$ excited states where calculated with high precision, resolving
even the smallest spin-dependent energy splittings. The calculation includes dynamical $u$, $d$, and $s$ sea quarks,
and was done at two different lattice spacings of approximately 0.11 fm and 0.08 fm. For lattice spacings of this order,
the $b$ quarks can be implemented accurately using lattice NRQCD \cite{Lepage:1992tx}. With lattice NRQCD it is also
possible to study individually the effects of the different spin-dependent interactions in the effective field theory
on the $bbb$ energy splittings.

\FloatBarrier
\section{Lattice actions}
\FloatBarrier

The lattice gauge-field configurations used in this work to perform the path integral were generated by the RBC/UKQCD
collaboration and are described in Ref.~\cite{Aoki:2010dy}. These configurations include the vacuum-polarization effects
of the light and strange quarks. The Iwasaki discretization was chosen for the gauge action, and the $u$, $d$, and $s$
quarks were implemented using a domain-wall action that preserves chiral symmetry even at nonzero lattice spacing.
The main parameters of the ensembles are given in Table \ref{tab:params}.

\vspace{6ex}

\begin{table}[h]
\begin{center}
\small
\begin{tabular}{cclcccl}
\hline \hline
$L^3 \times T$   & $\beta$ & $a m_{u,d}$ & $a m_s$ & $a m_b$ & $a\:\:[{\rm fm}]$ & $m_\pi\:\:[{\rm GeV}]$ \\
\hline
$24^3 \times 64$ & $2.13$  & $0.005$     & $0.04$  & $2.487$ & $0.1119(17)$      & $0.3377(54)$           \\ 
$24^3 \times 64$ & $2.13$  & $0.01$      & $0.04$  & $2.522$ & $0.1139(19)$      & $0.4194(70)$           \\ 
$24^3 \times 64$ & $2.13$  & $0.02$      & $0.04$  & $2.622$ & $0.1177(29)$      & $0.541(14)$            \\ 
$24^3 \times 64$ & $2.13$  & $0.03$      & $0.04$  & $2.691$ & $0.1196(29)$      & $0.641(15)$            \\ 
\\[-2ex]
$32^3 \times 64$ & $2.25$  & $0.004$     & $0.03$  & $1.831$ & $0.0849(12)$      & $0.2950(40)$           \\ 
$32^3 \times 64$ & $2.25$  & $0.006$     & $0.03$  & $1.829$ & $0.0848(17)$      & $0.3529(69)$           \\  
$32^3 \times 64$ & $2.25$  & $0.008$     & $0.03$  & $1.864$ & $0.0864(12)$      & $0.3950(55)$           \\ 
\hline \hline
\end{tabular}
\end{center}
\caption{\label{tab:params}The lattice sizes, the gauge couplings ($\beta=6/g^2$), the quark masses, and the
corresponding lattice spacings and pion masses.}
\end{table}

\newpage

In the following, we denote the $b$-quark field by $\psi$. Similarly to Ref.~\cite{Lepage:1992tx},
the lattice NRQCD action is written as
\begin{equation}
S_{\psi}=a^3\sum_{\mathbf{x},t}\psi^\dagger(\mathbf{x},t)\left[{\psi}(\mathbf{x},t)-\left(1-\frac{a\:\delta H}{2}\right)\!
\left(1-\frac{a H_0}{2n} \right)^{\!\!n} U_4^\dag \left(1-\frac{a H_0}{2n} \right)^{\!\!n}\!
\left(1-\frac{a\:\delta H}{2}\right)  {\psi}(\mathbf{x},t-a) \right], \label{eq:latact}
\end{equation}
where $U_4$ are the temporal gauge links, and $H_0$ and $\delta H$ are given by
\begin{eqnarray}
\nonumber H_0 &=& -\frac{\Delta^{(2)}}{2 m_b}, \label{eq:H0} \\
\nonumber\delta H&=&-c_1\:\frac{\left(\Delta^{(2)}\right)^2}{8 m_b^3}
+c_2\:\frac{ig}{8 m_b^2}\:\Big(\bs{\nabla}\cdot\mathbf{\widetilde{E}}
-\mathbf{\widetilde{E}}\cdot\bs{\nabla}\Big)-c_3\:\frac{g}{8 m_b^2}\:\bs{\sigma}\cdot
\left(\bs{\widetilde{\nabla}}\times\mathbf{\widetilde{E}}-\mathbf{\widetilde{E}}\times\bs{\widetilde{\nabla}} \right)
-c_4\:\frac{g}{2 m_b}\:\bs{\sigma}\cdot\mathbf{\widetilde{B}}\\
\nonumber&& + c_5\:\frac{a^2\Delta^{(4)}}{24m_b}-c_6\:\frac{a\left(\Delta^{(2)}\right)^2}{16n\:m_b^2}\\
\nonumber &&-c_7\:\frac{g}{8 m_b^3}\Big\{ \Delta^{(2)}, \: \bs{\sigma}\cdot\mathbf{\widetilde{B}} \Big \}
-c_8\:\frac{3g}{64 m_b^4}\left\{ \Delta^{(2)}, \: \bs{\sigma}\cdot \left(\bs{\widetilde{\nabla}}
\times\mathbf{\widetilde{E}}-\mathbf{\widetilde{E}}\times\bs{\widetilde{\nabla}} \right) \right\}
-c_9\:\frac{i g^2}{8 m_b^3}\:\bs{\sigma}\cdot(\mathbf{\widetilde{E}}\times\mathbf{\widetilde{E}}). \\
\label{eq:dH_full}
\end{eqnarray}
The different terms in the NRQCD action are suppressed by different powers of $v$, the average speed of the $b$ quarks
inside the hadron. The leading term, $H_0$, is of order $v^2$, and gives the dominant contribution to radial/orbital
energy splittings in the $b\bar{b}$ and $bbb$ systems. The terms with coefficients $c_{1,2,3,4}$ are the relativistic
corrections of order $v^4$, and the terms with coefficients $c_{7,8,9}$ are relativistic corrections of order $v^6$
(only the spin-dependent order-$v^6$ terms are included). The terms with coefficients $c_{5,6}$, which contain powers
of the lattice spacing, reduce discretization errors and are not present in the continuum NRQCD action. Because spin
splittings first arise at order $v^4$ through the operators $\bs{\sigma}\cdot\left(\bs{\widetilde{\nabla}}\times\mathbf{\widetilde{E}}
-\mathbf{\widetilde{E}}\times\bs{\widetilde{\nabla}} \right)$ and $\bs{\sigma}\cdot\mathbf{\widetilde{B}}$, the matching
coefficients $c_3$ and $c_4$ of these operators were tuned nonperturbatively to achieve high precision \cite{Meinel:2012qz}.
The other matching coefficients were set to their tree-level values, $c_i=1$.

\FloatBarrier
\section{Construction of $bbb$ operators}
\FloatBarrier

The $bbb$ energies can be extracted from Euclidean two-point functions of interpolating operators $\Omega^\Lambda_r$ with
the desired quantum numbers of the $bbb$ states on the lattice. These operators were constructed starting from
Gaussian-smeared $b$ quark fields
\begin{equation}
\tilde{\psi}_{{a}{\alpha}}= \left[\left(1+ \frac{r_S^2}{2 n_S}\Delta^{(2)} \right)^{n_S} \psi \right]_{{a}{\alpha}},
\end{equation}
where $a=1,2,3$ is the color index, $\alpha=\ua,\da$ is the spin index, $\Delta^{(2)}$ is a gauge-covariant lattice
Laplace operator, and the smearing radius was chosen to be $r_S\approx 0.14$ fm. To obtain a nontrivial spatial structure,
up to 2 gauge-covariant derivatives were then applied to obtain the following 13 quark building blocks,
\begin{eqnarray}
\nonumber \tilde{\psi}_{{a}{\alpha}{1}} &=& \tilde{\psi}_{{a}{\alpha}}, \\
\nonumber \tilde{\psi}_{{a}{\alpha}{2}} &=& (\nabla_x\: \tilde{\psi})_{{a}{\alpha}},
\hspace{2ex} \tilde{\psi}_{{a}{\alpha}{3}} = (\nabla_y\: \tilde{\psi})_{{a}{\alpha}},
\hspace{2ex} \tilde{\psi}_{{a}{\alpha}{4}} = (\nabla_z\: \tilde{\psi})_{{a}{\alpha}}, \\
 \tilde{\psi}_{{a}{\alpha}{5}} &=& (\nabla_x\: \nabla_x\: \tilde{\psi})_{{a}{\alpha}},
 \hspace{2ex} \tilde{\psi}_{{a}{\alpha}{6}} = (\nabla_y\: \nabla_x\: \tilde{\psi})_{{a}{\alpha}},
 \hspace{2ex} ...\:, \hspace{2ex} \tilde{\psi}_{{a}{\alpha}{13}} = (\nabla_z\: \nabla_z\: \tilde{\psi})_{{a}{\alpha}}.
 \label{eq:derivs}
\end{eqnarray}
Using Clebsch-Gordan coefficients, these building blocks were then combined to baryon operators $\Omega^J_m$ that would
have a definite total angular momentum $J$ in continuous space:
\begin{equation}
 \Omega^J_m =  \sum_{m_L, m_S} \langle J, m | L, m_L; S, m_S \rangle \:\: \Gamma_{{ijk\:\alpha\beta\gamma}}(L,m_L,S,m_S)
 \:\: \epsilon_{{abc}}\:\: \tilde{\psi}_{{a} {\alpha} {i}}\: \tilde{\psi}_{{b} {\beta} {j}}\: \tilde{\psi}_{{c} {\gamma} {k}}\,.
 \label{eq:Omegacont}
\end{equation}
There are multiple ways of combining the spin indices $\alpha,\beta,\gamma$ and the derivative indices $i,j,k$ to a given
total spin $S$ and total orbital angular momentum $L$, leading to different types of operators classified by permutation
symmetries \cite{Edwards:2011jj}. With two derivatives, the largest possible value of $J$ is $\frac72$.

Next, in order to account for the breaking of rotational symmetry by the lattice, linear combinations of the different
$m$-components of the operators $\Omega^J_m$ were formed, such that the resulting operators transform in irreducible
representations, $\Lambda$, of the double cover of the octahedral group $^2{\mathrm{O}}$:
\begin{equation}
\Omega^\Lambda_r = \sum_m \mathcal{S}^{J,m}_{\Lambda, r}\: \Omega^J_m. \label{eq:Omegalat}
\end{equation}
The subduction coefficients $\mathcal{S}^{J,m}_{\Lambda, r}$ can be found in Ref.~\cite{Edwards:2011jj}. The index
$r=1...{\rm dim}(\Lambda)$ denotes the row of the irrep. The full procedure described in this section produces seven
different $bbb$ operators in the $H_g$ irrep, three different operators each in the $G_{1g}$ and $G_{2g}$ irreps, and
one operator each in the $H_u$ and $G_{1u}$ irreps \cite{Meinel:2012qz}, where the subscripts $g/u$ denote even/odd parity.

\FloatBarrier
\section{Data analysis}
\FloatBarrier

Using the operators from Eq.~(\ref{eq:Omegalat}), the row-averaged two-point functions
\begin{equation}
 C_{ij}^{(\Lambda)}(t)=\frac{1}{{\rm dim}(\Lambda)}\sum_{r=1}^{{\rm dim}(\Lambda)}\big
 \langle \sum_\mathbf{x} \Omega^{\Lambda (i)}_r(\mathbf{x}, t) \: \Omega^{\Lambda (j)\dag}_r(0) \big\rangle \label{eq:twopt}
\end{equation}
were computed. Here, the indices $i,j$ label the different operators constructed in a given irrep $\Lambda$. For example,
in the $G_{1g}$ irrep, there are three different operators which were constructed using the following values of ($L$, $S$, $J$):
$(0,\frac12,\frac12)$, $(2,\frac32,\frac12)$, and $(2,\frac32,\frac72)$. The numerical results for (\ref{eq:twopt}) were
then fitted using the functions
\begin{equation}
C_{ij}^{\Lambda}(t) = \sum_{n=1}^N A_{n,i}^{(\Lambda)} A_{n,j}^{(\Lambda)} \:\: e^{-E_n^{(\Lambda)}\:t},
\end{equation}
where the number of exponentials, $N$, was chosen equal to the number of operators for each irrep. The full spectral decomposition
of $C_{ij}^{\Lambda}(t)$ also contains an infinite tower of higher excited states, but the starting value of $t$ was chosen
large enough such that the contributions from these states are negligible.

\begin{figure}[!ht]
\framebox{\includegraphics[width=0.395\linewidth]{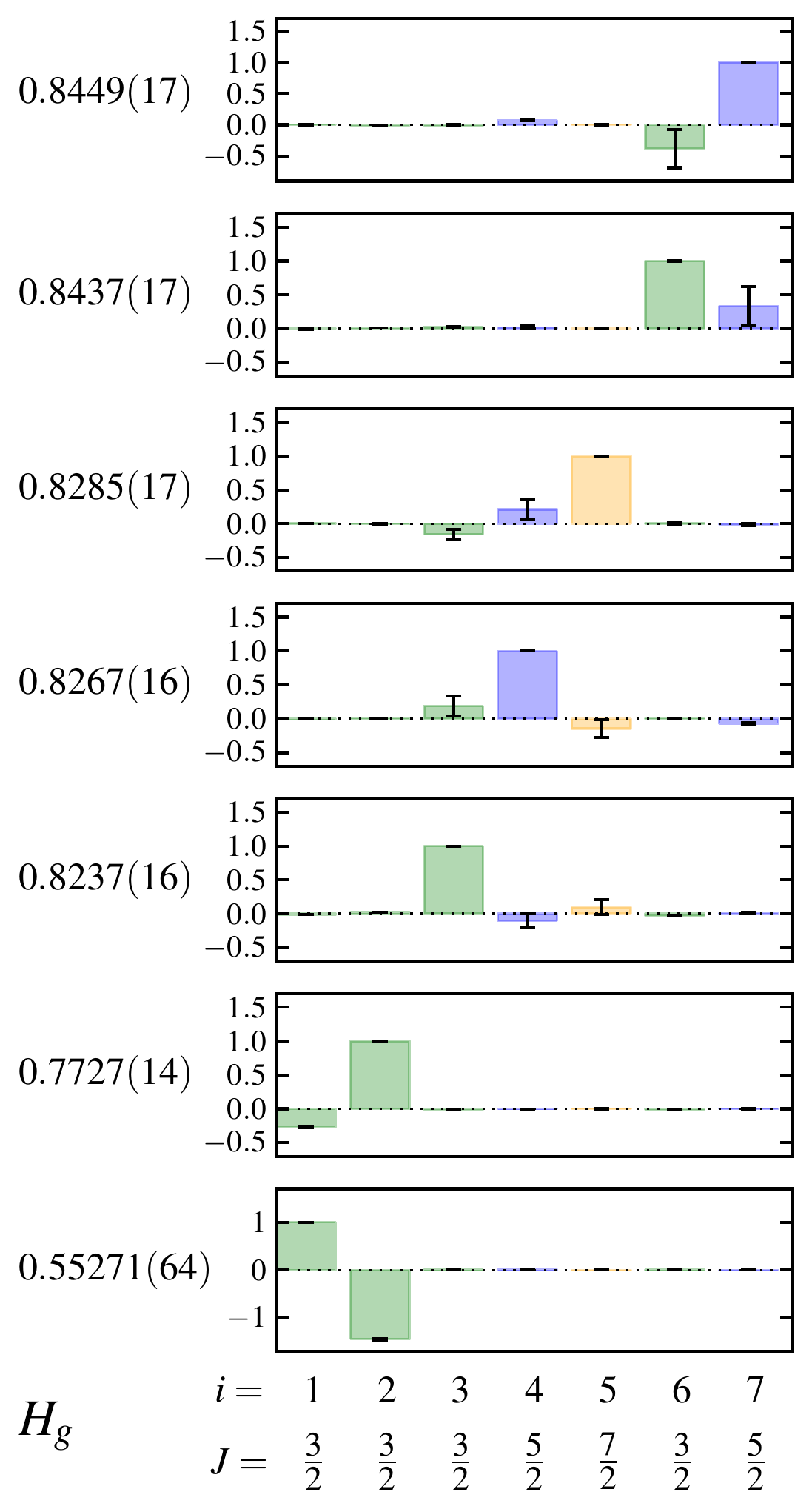}}
\hfill \framebox{\includegraphics[width=0.248\linewidth]{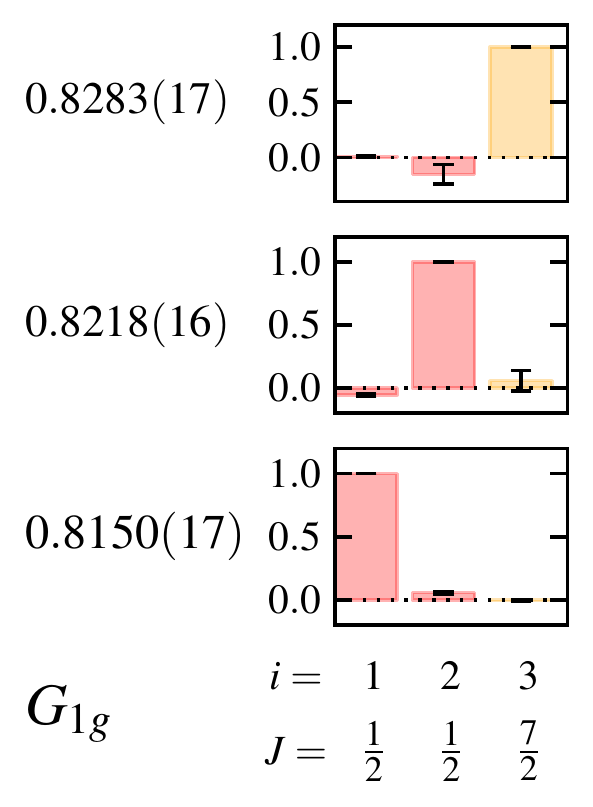}}
\hfill \framebox{\includegraphics[width=0.248\linewidth]{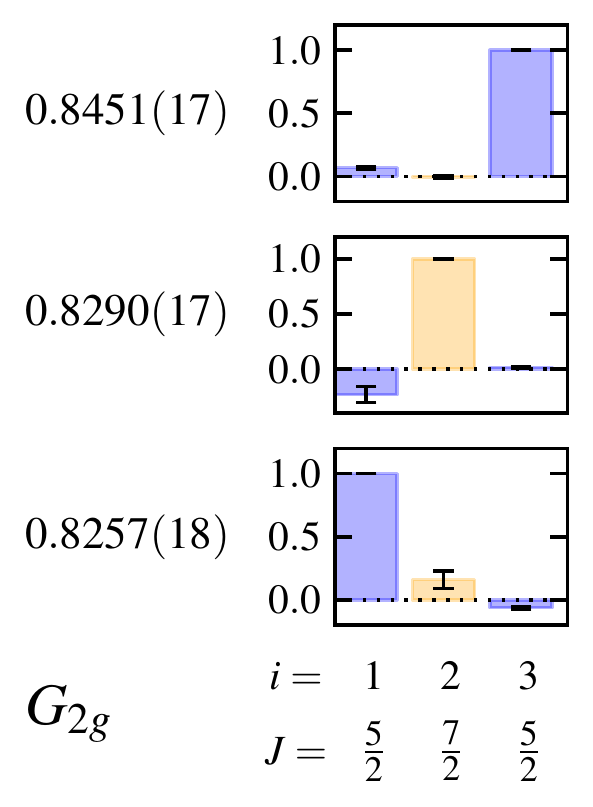}} 
\caption{\label{fig:spin_identification_L32_004_Hg} Fitted energies $aE_n^{(\Lambda)}$ and relative amplitudes
$A_{n,i}^{(\Lambda)}/A_{n,n}^{(\Lambda)}$ in the $H_g$, $G_{1g}$, and $G_{2g}$ irreps. The values of $J$ from
which the operators $\Omega^{\Lambda (i)}$ were subduced are indicated below the plots and by the colors.
The data shown here are for $a\approx 0.08$ fm, $a m_{u,d}=0.004$. }
\end{figure}

Sample fit results for $E_n^{(\Lambda)}$ and $A_{n,i}^{(\Lambda)}$ are shown in Fig.~\ref{fig:spin_identification_L32_004_Hg}.
The value of $J$ from which each operator $\Omega^{\Lambda (i)}$ was subduced is also shown. It turns out that for each
energy level $E_n^{(\Lambda)}$, the amplitudes $A_{n,i}^{(\Lambda)}$ are large only for one value of $J$, and this value
of $J$ can therefore be assigned to the energy level $E_n^{(\Lambda)}$. The operators retain a strong ``memory'' of the
original value of $J$ \cite{Edwards:2011jj}, and also $L$ and $S$ \cite{Meinel:2012qz}. Approximately matching energy
levels with $J=\frac52$ are seen to appear in both the $H$ and $G_2$ irreps, while approximately matching energy levels
with $J=\frac72$ appear in all three irreps, as expected.

\begin{figure}[!ht]
\includegraphics[height=0.22\textheight]{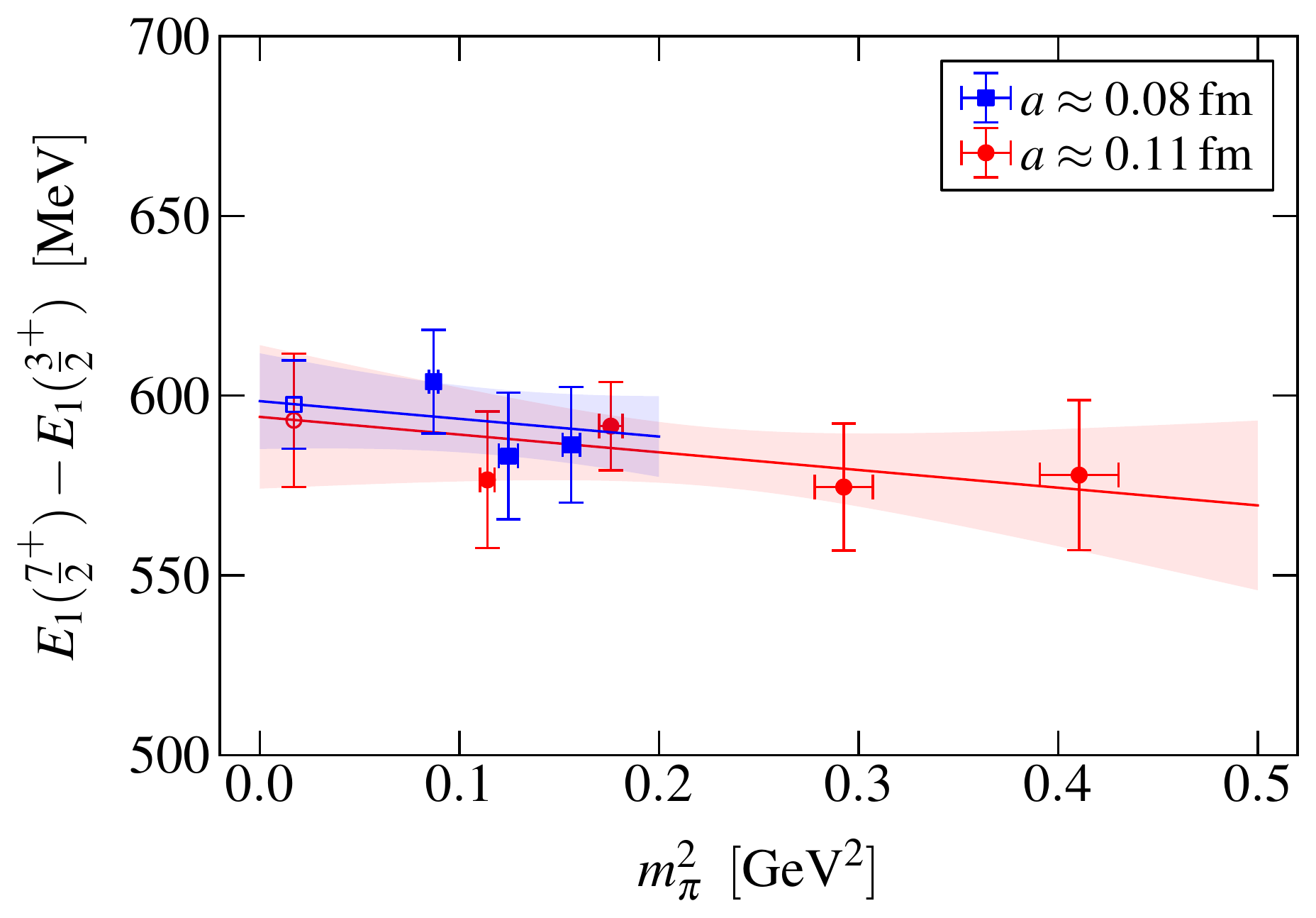} \hfill \includegraphics[height=0.22\textheight]{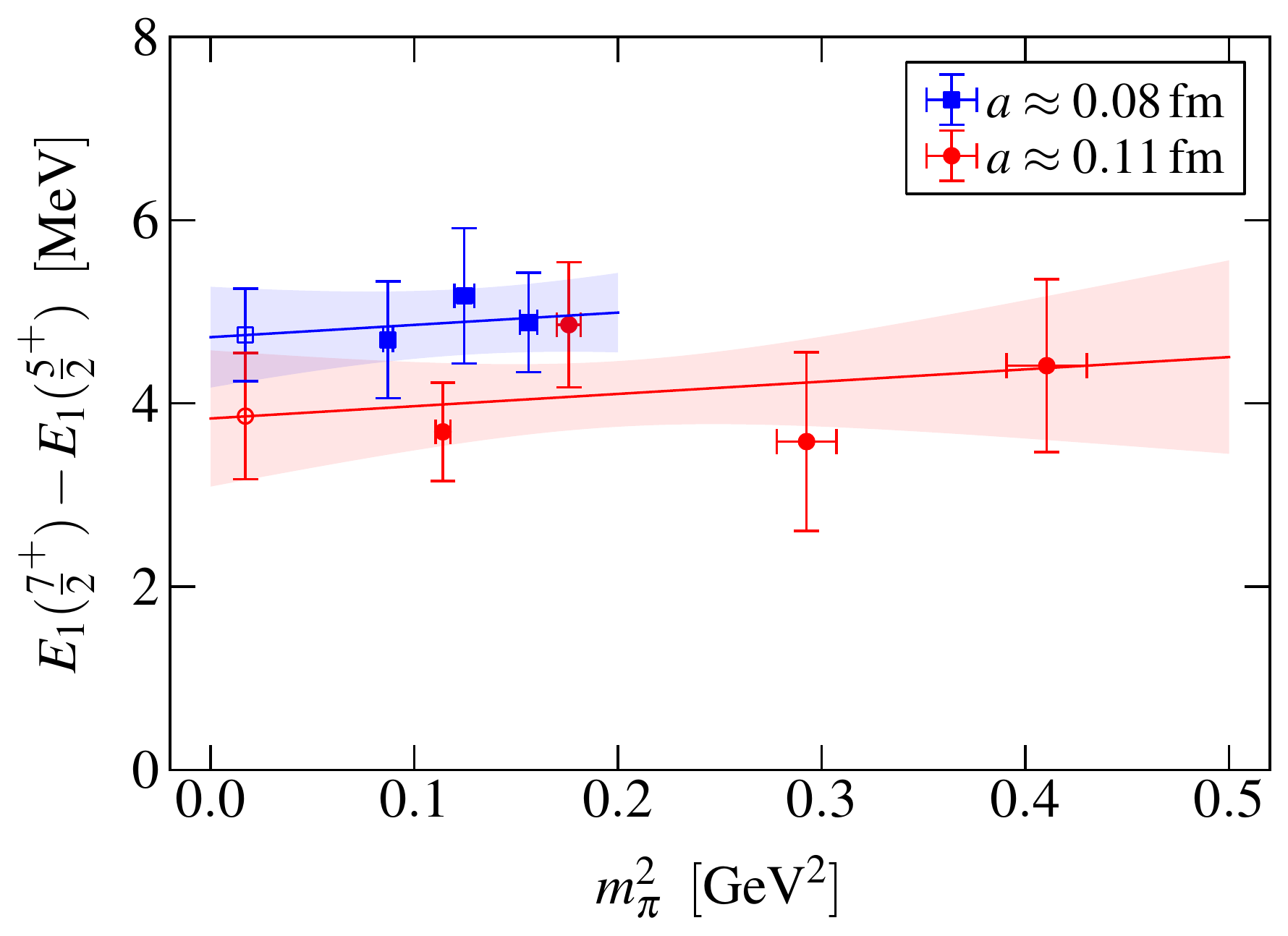}
\vspace{-2ex}
\caption{\label{fig:chiralextrap}Examples of chiral extrapolations:
$E_1(\frac72^+)-E_1(\frac32^+)$ (left) and $E_1(\frac72^+)-E_1(\frac52^+)$ (right).}
\vspace{-1ex}
\end{figure}

\noindent It is shown in Ref.~\cite{Meinel:2012qz} that the small splittings of matching energy levels
between the different irreps become smaller when the lattice spacing is reduced, demonstrating the restoration of
rotational symmetry. To get the best estimates of the continuum energy levels, simultaneous fits were performed across
the different irreps, using common energy parameters for the previously identified matching levels \cite{Meinel:2012qz}.
With the values of $J$ assigned, the energies can then be relabeled as $E_n(J^P)$ by

\begin{figure}[!t]
\vspace{-3ex}
\begin{center}
\includegraphics[width=0.78\linewidth]{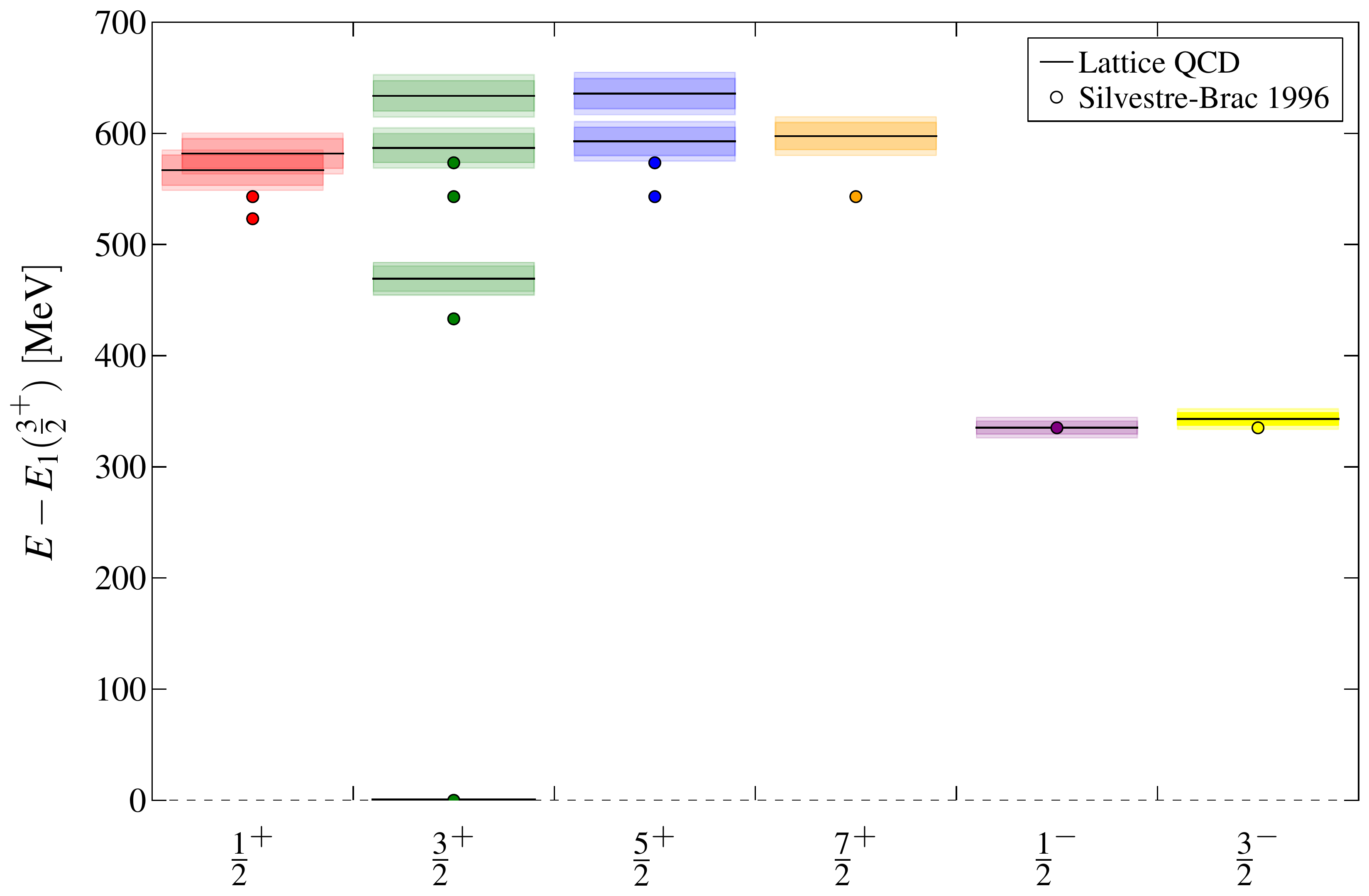}
\end{center}
\vspace{-4ex}
\caption{\label{fig:final}Final results for the $bbb$ energy splittings with respect to ground state $E_1(\frac 32^+)$,
compared to the potential-model calculation of Ref.~\cite{SilvestreBrac:1996bg}. For the lattice QCD results, the inner
shaded bands give the statistical/fitting/scale setting uncertainty, while the outer shaded bands also include the
systematic uncertainty. Because the results for the different energy levels are highly correlated, splittings between
nearby energy levels can be computed with much smaller absolute uncertainties (see Fig.~\protect\ref{fig:finalspin}).}
\end{figure}
\begin{figure}[!t]
\includegraphics[height=0.27\textheight]{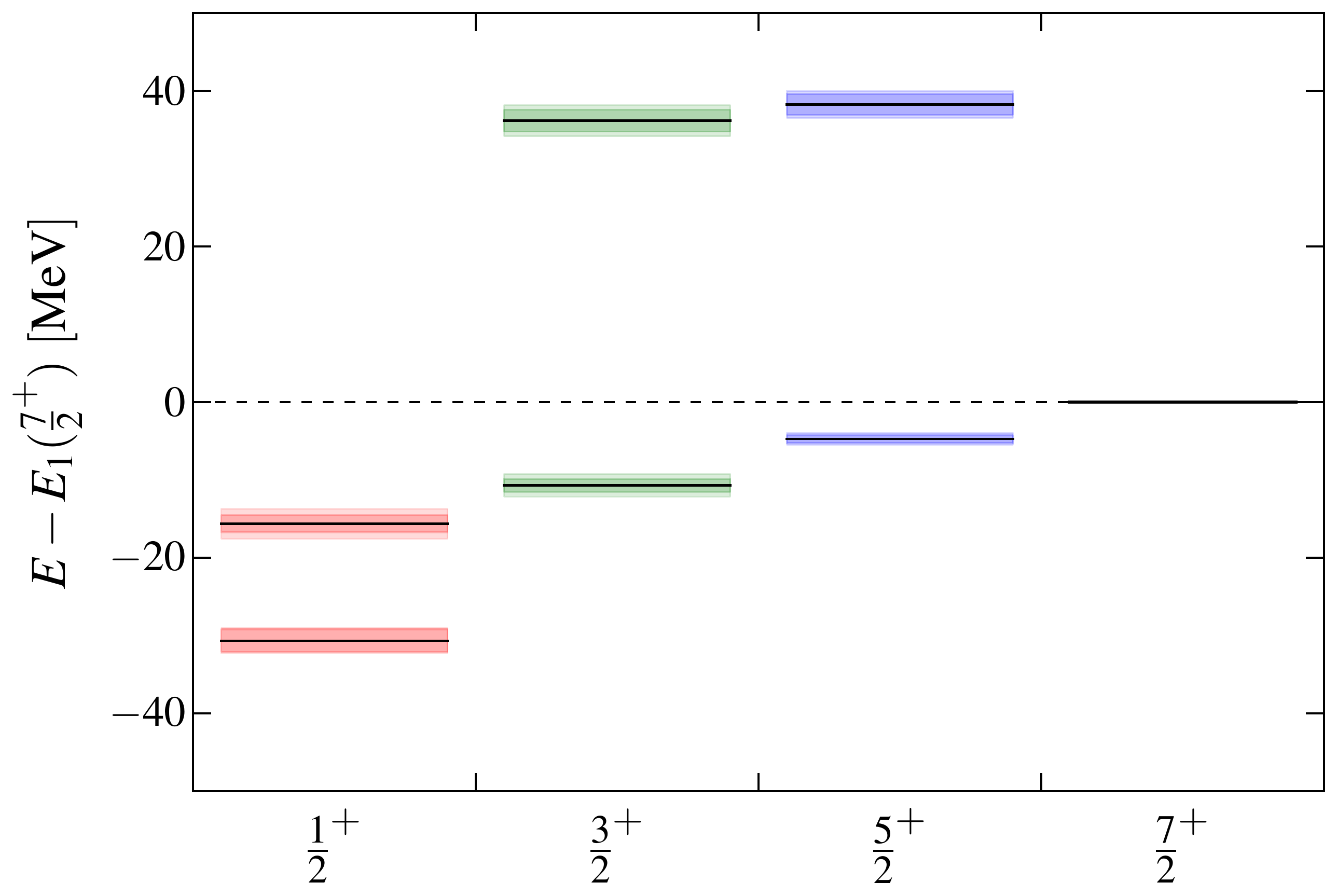}
\hfill \includegraphics[height=0.27\textheight]{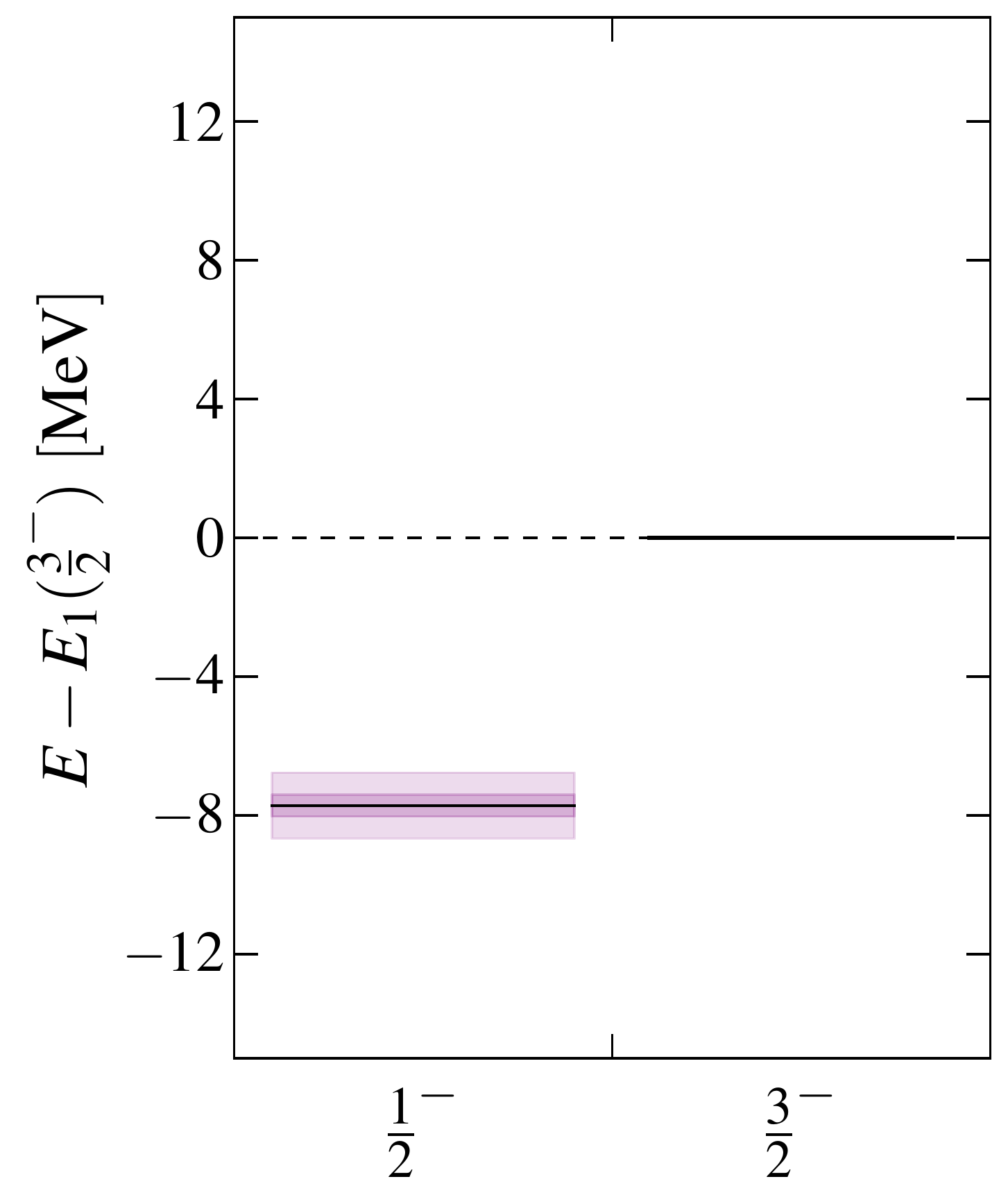}
\caption{\label{fig:finalspin}Final results for the $bbb$ energy splittings with respect to $E_1(\frac72^+)$ (left) and
$E_1(\frac32^-)$ (right). The inner shaded bands give the statistical/fitting/scale setting uncertainty, while the outer
shaded bands also include the systematic uncertainty.}
\end{figure}

\noindent increasing value for each $J^P$ channel. Note that, due to the use of NRQCD, only energy \emph{differences}
are physical. The energy differences computed on the various gauge-field ensembles were then extrapolated to the physical
value of the pion mass. These extrapolations were done simultaneously for the two different lattice spacings, as shown in
Fig.~\ref{fig:chiralextrap}. The results at the finer lattice spacing and at the physical pion mass can be taken as the
final values for the physical $bbb$ spectrum. The remaining systematic uncertainties were estimated individually for each
energy splitting in Ref.~\cite{Meinel:2012qz} by considering the contributions of the various NRQCD interaction terms.

In Fig.~\ref{fig:final}, the final results for the $bbb$ energy splittings with respect to the ground state are compared
to the potential-model calculation of Ref.~\cite{SilvestreBrac:1996bg}, where the potential is a sum of two-body
interactions and includes spin-spin interactions, but does not include spin-orbit or tensor interactions. For the gross
structure of the spectrum, the agreement of this model with the lattice QCD results is remarkably good. However, because
the model has no spin-orbit or tensor interactions, it predicts $E_2(\frac12^+)=E_3(\frac32^+)=E_1(\frac52^+)=E_1(\frac72^+)$,
\hspace{0.3ex} $E_4(\frac32^+)=E_2(\frac52^+)$, and $E_1(\frac12^-)=E_1(\frac32^-)$. As can be seen in Fig.~\ref{fig:finalspin},
all of these degeneracies are lifted in full QCD.

\FloatBarrier
\section{Contributions of individual NRQCD interactions}
\FloatBarrier

\begin{figure}[!hb]
\begin{center}
\includegraphics[width=\linewidth]{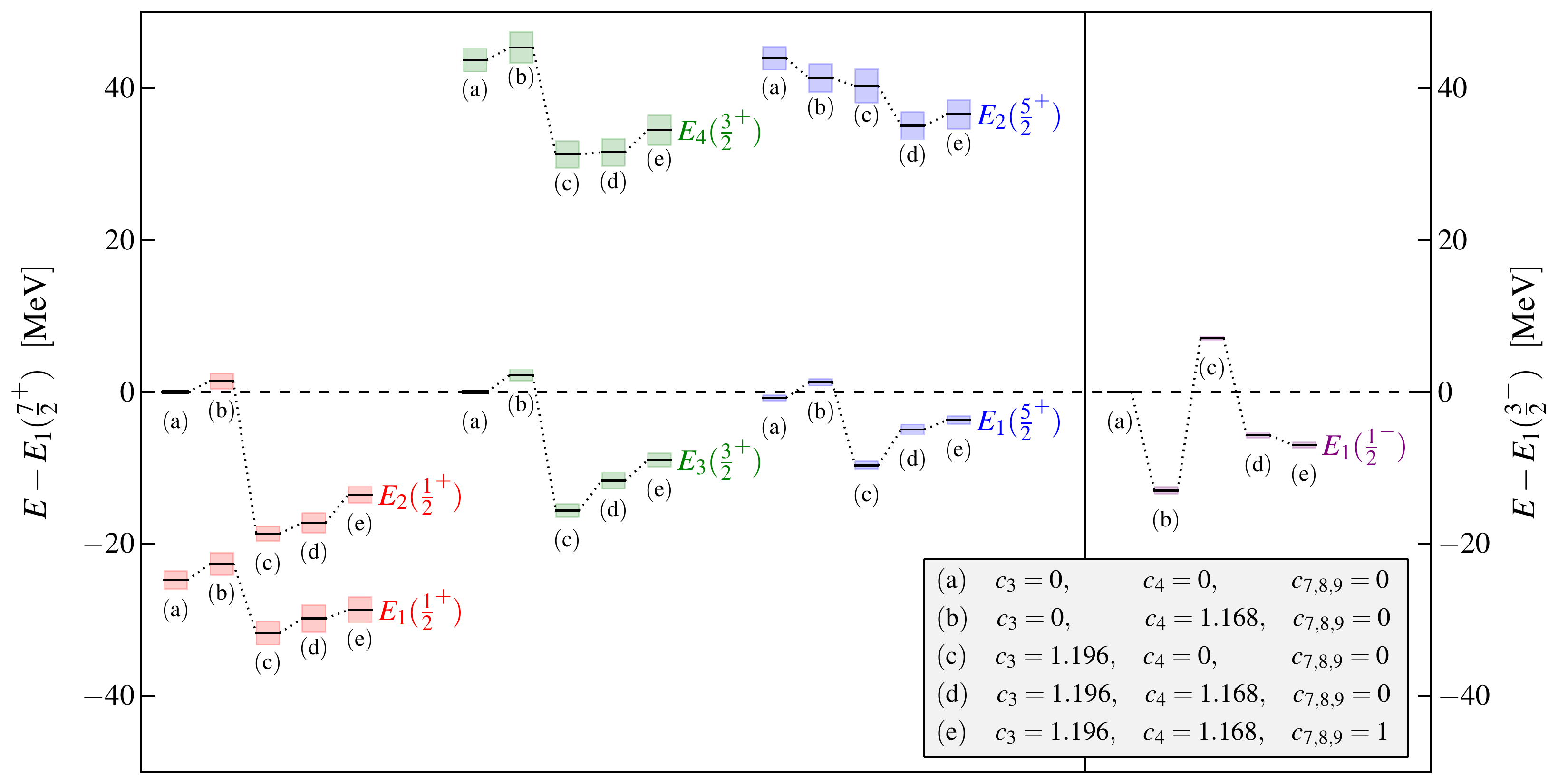} 
\end{center}
\caption{\label{fig:coeffdep}Dependence of the energy splittings $E_n(J^+) - E_1(\frac72^+)$ (left) and $E_1(\frac12^-)
- E_1(\frac32^-)$ (right) on the coefficients of the spin-dependent NRQCD interactions [see Eq.~(\protect\ref{eq:dH_full})].
For each energy level, five different choices of coefficients are shown. These results are from the ensemble with
$a\approx0.11$ fm, $a m_{u,d}=0.005$.}
\end{figure}

An interesting question is the following: what are the effects of the different spin-dependent operators in the NRQCD
action, Eq.~(\ref{eq:dH_full}), on the $bbb$ energy splittings? This was investigated in Ref.~\cite{Meinel:2012qz} on
one ensemble of gauge configurations ($a\approx0.11$ fm, $a m_{u,d}=0.005$) by setting different subsets of the NRQCD
coefficients, $c_i$, to zero, and recalculating the $bbb$ spectrum in each case. The results are shown here in
Fig.~\ref{fig:coeffdep}. In case (a), all spin-dependent interactions are turned off. Up to small corrections caused by
the breaking of rotational symmetry on the lattice, this leads to the degeneracies
$E_2(\frac12^+)=E_3(\frac32^+)=E_1(\frac52^+)=E_1(\frac72^+)$, \hspace{0.3ex} $E_4(\frac32^+)=E_2(\frac52^+)$,
and $E_1(\frac12^-)=E_1(\frac32^-)$, which are also present in the potential model of Ref.~\cite{SilvestreBrac:1996bg}.

In case (b), the operator $\bs{\sigma}\cdot\mathbf{\widetilde{B}}$ is turned on. Interestingly, the biggest effect is
seen for negative-parity energy levels, where this operator introduces a splitting of $E_1(\frac12^-)-E_1(\frac32^-)
= -12.97(45)$ MeV. Note that potential models with both spin-spin and tensor interactions, but without spin-orbit
interactions, predict $E_1(\frac12^-)-E_1(\frac32^-) = 0$ for $\Omega$ baryons \cite{Isgur:1978xj}, which suggests
that the splitting seen here is of spin-orbit type.

As shown in case (c), the operator $\bs{\sigma}\cdot\big(\bs{\widetilde{\nabla}}\times\mathbf{\widetilde{E}}
-\mathbf{\widetilde{E}}\times\bs{\widetilde{\nabla}} \big)$ has a large effect on both the positive-parity and negative-parity
$bbb$ energy splittings, which is of opposite sign to the effect of $\bs{\sigma}\cdot\mathbf{\widetilde{B}}$. For the multiplet
$\{ E_2(\frac12^+), E_3(\frac32^+), E_1(\frac52^+), E_1(\frac72^+)\}$, which in quark models has $L=2$, $S=\frac32$, and
a totally symmetric spatial wave function, the energy shifts due to $\bs{\sigma}\cdot\big(\bs{\widetilde{\nabla}}
\times\mathbf{\widetilde{E}}-\mathbf{\widetilde{E}}\times\bs{\widetilde{\nabla}} \big)$ are approximately proportional to
$\:2\,\mathbf{L}\cdot\mathbf{S}=J(J+1)-L(L+1)-S(S+1)$, which is characteristic for a spin-orbit interaction \cite{Gromes:1976cr}.

In case (d), both $\bs{\sigma}\cdot\mathbf{\widetilde{B}}$ and $\bs{\sigma}\cdot\big(\bs{\widetilde{\nabla}}
\times\mathbf{\widetilde{E}}-\mathbf{\widetilde{E}}\times\bs{\widetilde{\nabla}} \big)$ are turned on, and the resulting
shifts of the $bbb$ energy levels are approximately equal to the sum of the shifts from (b) and (c), but some deviations
from linearity can be resolved \cite{Meinel:2012qz}. Finally, in case (e), the spin-dependent order-$v^6$ interactions are
also included, restoring the full action as used in the main calculations of this work. The $bbb$ spin splittings change by
up to 30\% when these terms are included. A similarly large effect of the order-$v^6$ terms has also been observed for
the bottomonium spin splittings \cite{Meinel:2010pv}.

\FloatBarrier
\section{Conclusions}
\FloatBarrier

The $bbb$ system is an excellent benchmark for nonrelativistic effective field theories and potential models. Here, a
precise lattice QCD calculation of $bbb$ excited states up to $J=\frac72$ was presented. Spin-dependent energy splittings
that were neglected in potential models were resolved in this work, and it was shown what contributions these splittings
receive from the different NRQCD interactions. The author hopes that these results will stimulate further work on triply
heavy baryons using a variety of approaches.

\vspace{3ex}

\noindent \textbf{Acknowledgments:} This work is supported by the U.S.~Department of Energy under cooperative research
agreement Contract Number DE-FG02-94ER40818, and was also supported by the U.S.~Department of Energy under Grant Number
{D}{E}-{S}{C00}01{784}.


\begin{thebibliography}{99}

\bibitem{Brambilla:2005yk} 
  N.~Brambilla, A.~Vairo, and T.~Rosch,
  Phys.\ Rev.\ D {\bf 72}, 034021 (2005).

\bibitem{Brambilla:2009cd}
  N.~Brambilla, J.~Ghiglieri, and A.~Vairo,
  Phys.\ Rev.\  D {\bf 81}, 054031 (2010);
%
  F.~J.~Llanes-Estrada, O.~I.~Pavlova, and R.~Williams,
  Eur.\ Phys.\ J.\ C {\bf 72}, 2019 (2012);
%
  N.~Brambilla, F.~Karbstein, and A.~Vairo,
  arXiv:1301.3013.

\bibitem{Meinel:2010pw} 
  S.~Meinel,
  Phys.\ Rev.\ D {\bf 82}, 114514 (2010).

\bibitem{Meinel:2012qz} 
  S.~Meinel,
  Phys.\ Rev.\ D {\bf 85}, 114510 (2012).

\bibitem{Lepage:1992tx}
  G.~P.~Lepage {\it et al.},
  Phys.\ Rev.\  D {\bf 46}, 4052 (1992).
  
\bibitem{Aoki:2010dy} 
  Y.~Aoki {\it et al.} (RBC and UKQCD Collaborations),
  Phys.\ Rev.\ D {\bf 83}, 074508 (2011).
  
\bibitem{Edwards:2011jj} 
  R.~G.~Edwards, J.~J.~Dudek, D.~G.~Richards, and S.~J.~Wallace,
  Phys.\ Rev.\ D {\bf 84}, 074508 (2011).

\bibitem{SilvestreBrac:1996bg}
  B.~Silvestre-Brac,
  Few-Body Syst.\  {\bf 20}, 1 (1996).

\bibitem{Isgur:1978xj} 
  N.~Isgur and G.~Karl,
  Phys.\ Rev.\ D {\bf 18}, 4187 (1978);
%
  K.-T.~Chao, N.~Isgur, and G.~Karl,
  Phys.\ Rev.\ D {\bf 23}, 155 (1981);
%
  D.~Gromes,
  Z.\ Phys.\ C {\bf 18}, 249 (1983).

\bibitem{Gromes:1976cr} 
  D.~Gromes and I.~O.~Stamatescu,
  Nucl.\ Phys.\ B {\bf 112}, 213 (1976).

\bibitem{Meinel:2010pv} 
  S.~Meinel,
  Phys.\ Rev.\ D {\bf 82}, 114502 (2010).

\end{thebibliography}
\end{document}